\documentclass[preprint,aps,superscriptaddress,nofootinbib,showkeys,11pt]{revtex4}
\pdfoutput=1
\usepackage{graphicx}
\usepackage{amsmath}
\usepackage{amsfonts}
\usepackage{amssymb}
\usepackage{xcolor, soul}
\usepackage{epstopdf}
\usepackage{float}
\usepackage{subfigure}
\usepackage{hyperref}
\hypersetup{
     colorlinks   = true,
     citecolor    = red,
     linkcolor    = blue,
     urlcolor     = blue,
}
\def\beq{\begin{equation}}
\def\eeq{\end{equation}}
\def\bea{\begin{eqnarray}}
\def\eea{\end{eqnarray}}
\def\be{\begin{equation}}
\def\ee{\end{equation}}

\def\bse{\begin{subequations}}
\def\ese{\end{subequations}}

\graphicspath{{./figs/}}
\begin{document}

\title{Plasmon induced transparency in an air-dielectric grooved parallel plate terahertz waveguide}

\author{KM Dhriti}%
\email{dhrit176121006@iitg.ac.in}
\affiliation{%
	Department of Physics, Indian Institute of Technology Guwahati, Guwahati– 781039, Assam, India 
}%
\author{Maidul Islam}
\email{maidul.alig@gmail.com}
\affiliation{Department of Physics, The Assam Royal Global University, Guwahati, Assam - 781035, India}
\author{Angana Bhattacharya}%
\email{angana18@iitg.ac.in}
\affiliation{%
	Department of Physics, Indian Institute of Technology Guwahati, Guwahati– 781039, Assam, India
}%
\author{Amir Ahmad}%
\email{amirahmad01@gmail.com}
\affiliation{College of Information Technology, United Arab Emirates University, Al Ain, UAE}
\author{Gagan Kumar}%
\email{gaganphy.iitd@gmail.com}
\affiliation{%
	Department of Physics, Indian Institute of Technology Guwahati, Guwahati– 781039, Assam, India
}%

\date{\today}
\begin{abstract}
In this article, we examine plasmon induced transparency (PIT) effect in a parallel plate waveguide (PPWG) comprising of two similar pyramidal shaped grooves. One of the grooves is filled with air, while the other is filled with a dielectric material whose refractive index can be varied. The resonant frequencies corresponding to the air and dielectric grooves in the proposed configuration results in the transparency window which can be modulated with the refractive index of the dielectric material. The approach provides flexibility to actively control transparency effect in a waveguide configuration without changing physical dimensions. We examined field profiles in the transparency region to clearly depict the PIT effect. We have employed an analytical model based upon the three-level plasmonic model to validate our numerical findings. Further, we examined the switching and tunability of transparency effect by including silicon layers between the grooves whose conductivity can be varied. Actively tunable response in plasmon induced transparency effect in terahertz waveguides can be significant in the construction of terahertz active components.
\end{abstract}
\maketitle
\section{Introduction}
The terahertz (THz) band (0.1-10 THz) spans over a narrow region between the microwave and infrared regions in the electromagnetic spectrum. It exhibits nonionizing character because of its low energies \cite{1,2}. In last decade, significant attention has been paid to terahertz (THz) wave propagation due to its numerous practical applications in the field of sensing \cite{3}, medical imaging \cite{4}, spectroscopy \cite{5}, characterization of dielectric materials \cite{6}, pharmaceutical drug testing \cite{7}, astronomy \cite{8}, and communication \cite{8}. The atmosphere has a strong impact on the propagation of terahertz waves. Therefore, it is necessary to design guided transmission media for low loss, flexible, long distant and efficient transmission of terahertz waves. In this context, various types of guided transmission media have been proposed in the last decade. Metallic wires and metal coated dielectric tubes were proposed earlier but disregarded due to their higher bending loss and low coupling efficiency \cite{9, 10}. Especially, THz on-chip integrated circuits have drawn more attention owing to the potential applications in communication, imaging and chip scale sensing \cite{11,12,13}. In this direction, Wang et al. experimentally demonstrated how a simple waveguide like a metal wire has the ability to transport terahertz signals with low attenuation \cite{14}. Recently, several waveguide geometries have been examined both experimentally and numerically, such as polymer waveguides \cite{15}, silicon waveguides \cite{16} and metallic waveguides \cite{17}. However, metallic waveguide configurations viz. parallel plate waveguides, a waveguide comprising periodically arranged plasmonic structures have been widely explored as they exhibit negligible ohmic losses and strong field confinement. In the context of parallel plate configuration, Mendis et al. experimentally observed a metallic parallel plate waveguide to realize the filtering functionality by using artificial dielectric material \cite{18, 19}. Further, a parallel-plate waveguide with T-junction has been studied for propagating THz wave power transfer from a single waveguide channel into two different waveguide channels \cite{20}. To achieve highly confined terahertz surface modes researchers have developed periodically arranged plasmonic structures-based waveguide configuration. In this direction, a significant amount of work has been done with different shapes and sizes of the metallic corrugations. Zhu et al. investigated a thin metal sheet of stainless steel comprising of periodically patterned one-dimensional array of apertures, which supports highly confined terahertz surface modes \cite{21}. The strong confinement of surface modes has potential applications in a wide range of applications such as sensing \cite{22,23,24,DM}, control on propagation \cite{25,26,27}, slow light \cite{28, 29}, concentration of fields, and other active as well passive device constructions \cite{30, 31}. In spite of significant interest in terahertz waveguides, limited attention has been paid to explore plasmon induced transparency effect in terahertz guided systems and to actively control it, which is very important for the construction of futuristic active and passive components.\\ 
The destructive interference of two optical signals in a three atomic level system leads to the electromagnetically induced transparency (EIT) phenomenon \cite{32,33,34,35,36}. It has attracted a lot of attention because of its potential usage in various fields such as nonlinearities \cite{37}, optical data storage \cite{38}, modulations \cite{39}, ultra-high sensors \cite{40} and slow light application \cite{41, 42}. EIT effect has been realized in various configurations including coupled resonator systems \cite{43, 44}, metamaterials \cite{40, 45}, grating \cite{46}, and waveguides \cite{47,48,49,KM}. In waveguide configurations, Xu et al. have experimentally observed a silicon micro-ring resonator coupled to a parallel waveguide to realize transparency window by constructive interference \cite{50}. Recently, plasmon-induced transparency (PIT), an analogue of electromagnetically induced transparency (EIT), has drawn more attention due to the promising on-chip applications. In this context, Zhao et al. have demonstrated PIT effect in subwavelength metal structure waveguide consisting of metallic cut wires and double-gap split ring resonators \cite{51}. The investigations so far in this area have focused on passively tuning the PIT response. Actively tunable PIT effect, despite being significant, has not been examined so far to the best of our knowledge. There is a strong need to pursue research in this direction.\\ 
In this paper, we propose a metal-air-metal waveguide comprising double pyramidal groove structures, which exhibits the plasmon induced transparency (PIT) effect through destructive interference of two bright resonators. To study the active modulation of plasmon induced transparency window, we varied refractive index of the dielectric in one of the pyramidal grooves. The paper is organized as follows: first, we discuss the proposed waveguide geometry comprising two pyramidal grooves filled with air and dielectric. After that, we numerically examine the transmission spectra of the proposed waveguide for different cases viz. air and dielectric filled grooves only and both the grooves in parallel configuration. Next, we employ a theoretical model based on the three-level plasmonic model to understand and validate our numerically obtained transmission properties. After that, we examine electric field profiles to ensure PIT-effect is supported by the proposed waveguide. Further, we vary refractive index of the dielectric material to actively modulate the PIT window. Next, active modulation and switching of PIT effect are examined by placing a silicon sheet on the top of a groove and varying its conductivity. A comprehensive picture of active modulation of the PIT window with conductivity is presented through a contour plot. Finally, we summarize the results in the conclusion section.

\section{Schematic of 3D Waveguide}
\begin{figure}[h!]
\centering\includegraphics[width=11cm]{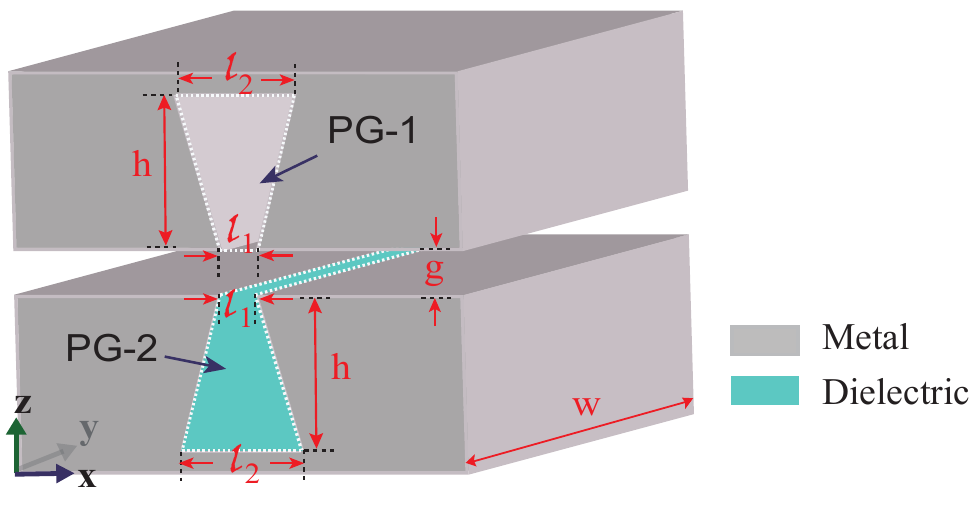}
\caption{Schematic of proposed metal-air-metal waveguide geometry: 3-D view of proposed waveguide design consisting of two parallel metallic blocks. Two pyramidal groove structures are designed in the two metallic blocks i.e. PG-1 and PG-2. The gap (g) between two blocks is fixed at $g = 100 \mu m$. The lower groove (PG-2) is filled with dielectric. The geometrical parameters of the grooves are as follows: $l_1 = 50 \mu m$, $l_2 = 120 \mu m$, $w = 800 \mu m$, $h = 200 \mu m$.}\label{fig1}
\end{figure}

The schematic illustration of the proposed metal-air-metal waveguide is shown in Fig.\ref{fig1}. The metal-air-metal structure consists of an air medium sandwiched between the two parallel metallic blocks. Two pyramidal groove structures are designed in the blocks along transverse direction. A gap of 100 um  is maintained between the two blocks.  One of the two pyramidal grooves is filled with a dielectric material of varying refractive index to study the active modulation of PIT window and the other one remains empty. We have used the terms PG-1 and PG-2 for pyramidal resonating grooves without and with dielectric, respectively, throughout the study. In the schematic, the different structure parameters of PG are denoted as: width ($w$), length ($l_1$ and $l_2$), depth ($h$), and gap between the two grooves ($g$). The following values of different parameters remain fixed throughout the study: $w = 800 \mu m$, $l_1 = 50 \mu m$, $l_2 = 120 \mu m$, $h = 200 \mu m$, $g = 100 \mu m$.\\ In our designed metal-air-metal waveguide comprising pyramidal grooves, a terahertz broadband signal is coupled to the one end of the waveguide. While propagating along x-axis in the air medium between the two parallel metallic slabs, it further couples to the pyramidal groove structures. Finally, it is detected at the other end of the waveguide. It is important to highlight that the proposed waveguide can be fabricated via conventional photolithography technique by using a highly doped silicon substrate while taking advantage of its crystalline structure. It is important to highlight that the proposed waveguide can be fabricated via conventional photolithography technique by using a highly doped silicon substrate while taking advantage of its crystalline structure. The corrugations of the waveguide depend on the crystal orientation of the silicon substrate. To make pyramidal grooves, one can use a crystalline silicon wafer of (100) orientation with dopant concentration, $n\geq 10^{19} cm^{-3}$ which behaves like a perfect conductor at terahertz frequencies. Using low-pressure chemical vapor deposition (LPCVD) technique, the silicon dioxide layer can be grown on the silicon surface and further, patterns can be made via photolithography technique \cite{b,c}. In the next step, appropriately patterned silicon can be etched in a mixture of potassium hydroxide, water, and isopropanol in the ratio of 60:30:10 to make inverted pyramidal apertures \cite{d}. For the fabrication of groove, a silicon wafer can be glued on the back of the pyramidal apertures using a conducting epoxy. For characterization, one can use the technique of terahertz time-domain spectroscopy (THz-TDS) \cite{52}. 
\section{Plasmon Induced Transparency in the Proposed Waveguide: Simulation and Theory}
\begin{figure}[h!]
\centering\includegraphics[width=13cm]{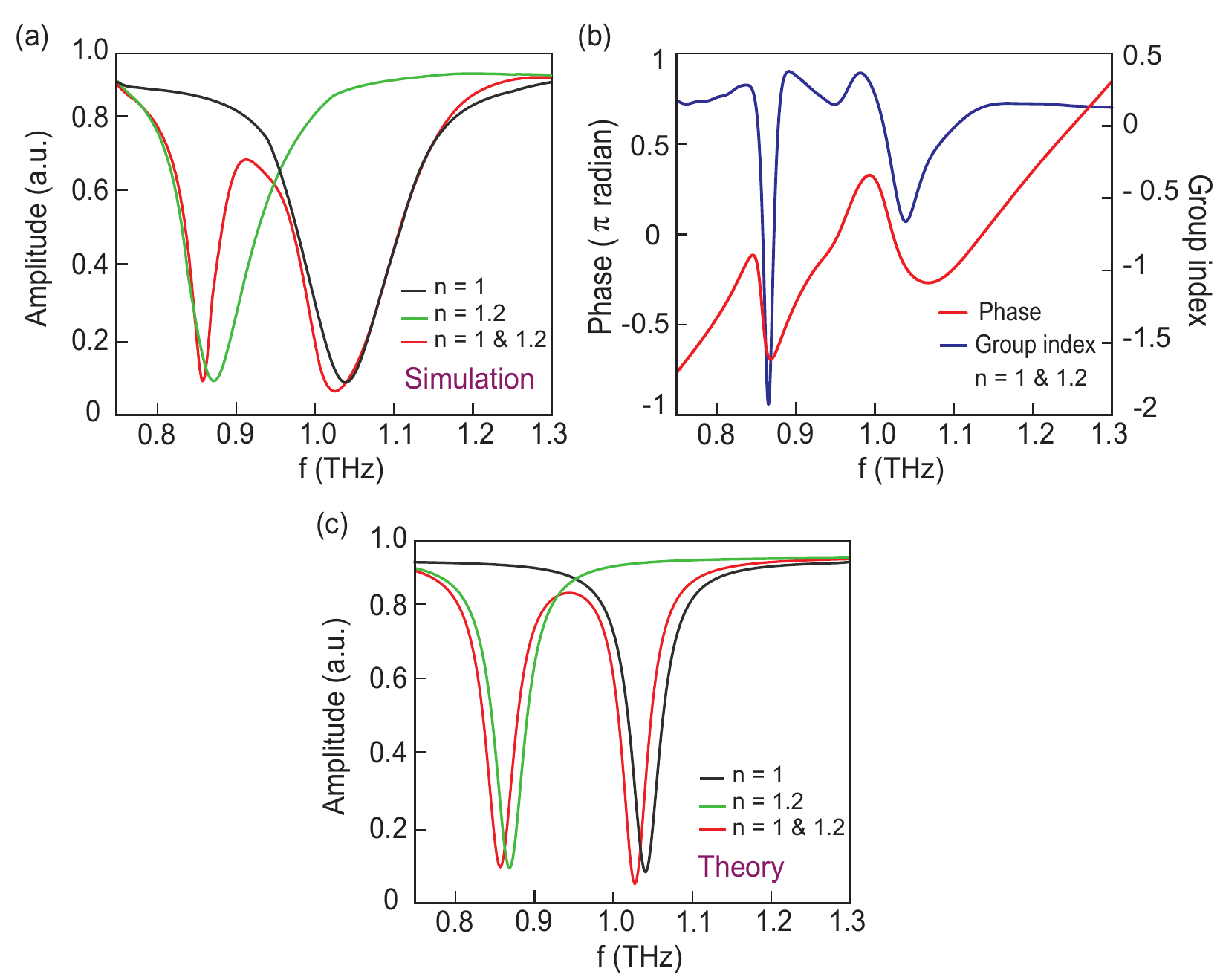}
\caption{(a) THz transmission spectra of proposed terahertz waveguide from simulation. Black, red, and blue traces correspond to PG-1, PG-2 and the combined structure, PG-1 $\&$ PG-2, respectively. (b) Numerically calculated phase and group index values of the transmitted terahertz versus frequency in the proposed waveguide geometry with dielectric material of n = 1.2 filled in PG-2. (c) Transmission characteristics from theoretical modelling based on three level plasmonic system. The different parameters of the proposed terahertz waveguide are kept fixed: $l_1 = 50 \mu m$, $l_2 = 120 \mu m$, $w = 800 \mu m$, $h = 200 \mu m$.}\label{fig2}
\end{figure}

We have performed the simulations with the help of finite element time domain solver in a commercially available Computer Simulation Technology (CST) microwave studio software to obtain the transmission properties. In simulations, open boundary conditions are assumed in all directions and metal is considered as a perfect electrical conductor (PEC) due to its high conductivity at terahertz frequencies. Single cycle z-polarized terahertz beam is incident at one end of the waveguide and allowed to propagate along the gap between two parallel metallic blocks. While propagating through the gap it couples to PG-1 $\&$ PG-2 and finally, it is probed at the receiver end of the waveguide in the form of time domain signal, which is converted into frequency domain using Fast Fourier Transform (FFT). First, we examine a waveguide design comprising PG-1 and PG-2 individually, which correspond to empty (air) and filled pyramidal groove of refractive index ‘n’ = 1 and 1.2, respectively. Next, we investigate the waveguide comprising both PG-1 and PG-2. Fig.\ref{fig2}(a) depicts the results of numerically calculated transmission properties of the proposed terahertz waveguide for PG-1, PG-2 individually and for both PG-1 and PG-2. In the figure, black, green and red traces correspond to numerically obtained transmission amplitude of the terahertz waveguide for PG-1, PG-2 and both PG-1 and PG-2, respectively. It can be noticed from the figure that PG-1 supports bright mode with resonance frequency $\omega_1$ = 1.04 THz, whereas PG-2 exhibits bright mode at $\omega_2$ = 0.87 THz. Hence, the waveguide supports two terahertz bright modes from the resonators of the waveguide at two distinct frequencies ($\omega_1$ and $\omega_2$) close to each other. These two bright modes experience destructive interference while propagating through the gap between the two parallel blocks of waveguide comprising both PG-1 and PG-2 and thereby exhibit plasmon induced transparency window at 0.92 THz.

After that, we examine the change of phase value of the transmitted terahertz amplitude. The numerically calculated phase value versus frequency is shown in Fig. 1(b) in red trace. A dramatic change in the phase value further elucidates the PIT effect. Further, we have calculated group index value corresponding to the phase plot, which is depicted in the same figure in the blue trace. The group index ($n_g$) is an important feature of the PIT effect and is associated with the phase by the following relation \cite{li},
\begin{equation*}
\label{eq6}
n_g=-\frac{c}{d} \frac{d\phi}{d\omega}
\end{equation*}
where, c is the velocity of light in free space, d is the total length of waveguide and $\phi$ is the phase of the transmitted terahertz. A sharp change in the group index signifies a highly dispersive character of the PIT effect which is important to several applications including slow light systems.
In order to validate our numerically obtained transmission properties and get a physical insight into the coupling mechanism of two bright modes involved in the plasmon induced transparency, we employ a three-level plasmonic model \cite{53}. In our case, waveguide supports two modes from the resonators of the waveguide at two different nearby frequencies. These modes are coupled via a strong electric field in near field configuration. The two resonant modes are termed as bright modes due to simultaneous excitation under the effect of incident terahertz beam. The amplitudes of these two radiative resonant modes can be expressed in terms of coupled Lorentz oscillators as
\begin{gather}
\begin{pmatrix}
\tilde{a}\\
\tilde{b}
\end{pmatrix}
=\frac{1}{\left(\delta+i\gamma_a\right)\left(\delta+i\gamma_b\right)-\kappa^2}
\begin{pmatrix}
\left(\delta + i\gamma_b\right) & -\kappa \\
-\kappa & \left(\delta+i\gamma_a\right)
\end{pmatrix}
\begin{pmatrix}
-G E_0 \\
-G E_0
\end{pmatrix}
\label{eq1}
\end{gather}
where $\tilde{a}$ and $\tilde{b}$ are induced THz field amplitude corresponding to the two resonators i.e. PG-1 and PG-2 respectively.  The two modes have adjacent resonance frequencies $\omega_1$ and $\omega_2$, such that $\delta=\omega-\omega_1=\omega-\omega_2$ is very small in Eq.\eqref{eq1}. $\gamma_a$  and $\gamma_b$ are the damping factor of the resonant modes, which are much smaller than the resonance frequencies. $E_0$  is the amplitude of the incident THz electric field whereas is the corresponding frequency of the incident THz signal;  $\kappa$ is the coupling coefficient between the two bright modes; $G$ is a geometric parameter which represents the coupling between the resonators and the incident THz electric field.\\
From Eq.\eqref{eq1}, the field amplitude of the $1^{st}$ mode can be expressed as, 
\begin{equation}
\label{eq2}
\tilde{a}=\frac{-GE_0\left(\delta + i\gamma_b\right)+{\kappa}GE_0}{\left(\delta+i\gamma_a\right)\left(\delta+i\gamma_b\right)-\kappa^2}\\
=\frac{GE_0\left[\kappa -\left(\omega-\omega_2 + i\gamma_b\right)\right]}{\left(\omega-\omega_1 + i\gamma_a\right)\left(\omega-\omega_2 + i\gamma_b\right)-\kappa^2}
\end{equation}
Similarly, for the $2^{nd}$ mode, it is given by,
\begin{equation}
\label{eq3}
\tilde{b}=\frac{GE_0\left[\kappa -\left(\omega-\omega_1 + i\gamma_a\right)\right]}{\left(\omega-\omega_1 + i\gamma_a\right)\left(\omega-\omega_2 + i\gamma_b\right)-\kappa^2}
\end{equation}
The transmission coefficient of the plasmonic waveguide can be written as:
\begin{equation*}
T=\left|\frac{\tilde{a}}{E_0}\right|^2+\left|\frac{\tilde{b}}{E_0}\right|^2
\end{equation*}	
\begin{equation}
\label{eq4}
T=|a|^2+|b|^2
\end{equation}
We use Eq.\eqref{eq1} to calculate the terahertz transmission coefficients separately for three different waveguide configurations comprising PG-1 only, PG-2 only and the combined structure of PG-1 $\&$ PG-2. The results of corresponding transmission spectra are shown in Fig.\ref{fig2}(c). Different color traces indicate transmission for the corresponding refractive index $‘n’$ value of different waveguide configurations. Theoretical modeling gives rise to synonymous transmission spectrum as obtained through numerical simulations for a specific set of values of the modeling parameters given in Table-\ref{tab:table1}. From the figures, it can be noticed that theoretically obtained dip frequencies of two resonant modes along with the plasmon induced window match with the numerically obtained results.
\begin{table}[ht]
	\centering
			\begin{tabular}{|c|c|c|c|c|c|}
			\hline
Resonator type& $G$&$\kappa$& $\gamma_a$& $\gamma_b$\\ &(THz)&(THz)&(THz)&(THz)\\
				\hline
			PG-1 & 0.23 & 0& 0.127&0.127  \\
			PG-2 & 0.23 & 0& 0.131&0.131 \\
			PG-1 $\&$ PG-2&0.3&0.04&0.112&0.125 \\
			\hline
			\end{tabular}
\caption{\label{tab:table1}Specific values of parameters for the theoretical modeling of PIT effect}		
\end{table}
From Table-\ref{tab:table1}, one can observe that the value of the coupling coefficient parameter is zero for PG-1 and PG-2 individually, since the resonators are excited separately i.e. in the absence of any other aperture resonator. Furthermore, it can be observed that the value of the geometric parameter increases when waveguide configuration is comprised of both the resonators PG-1 and PG-2 rather than a waveguide comprising of either of two resonators. This is due to the strong coupling of the two resonant modes supported by the combined structure of both PG-1 and PG-2 with the incident terahertz beam rather than the coupling of either of two resonant modes supported by either PG-1 or PG-2 in the proposed waveguide configurations.
\begin{figure}[h!]
\centering\includegraphics[width=13cm,height=3cm]{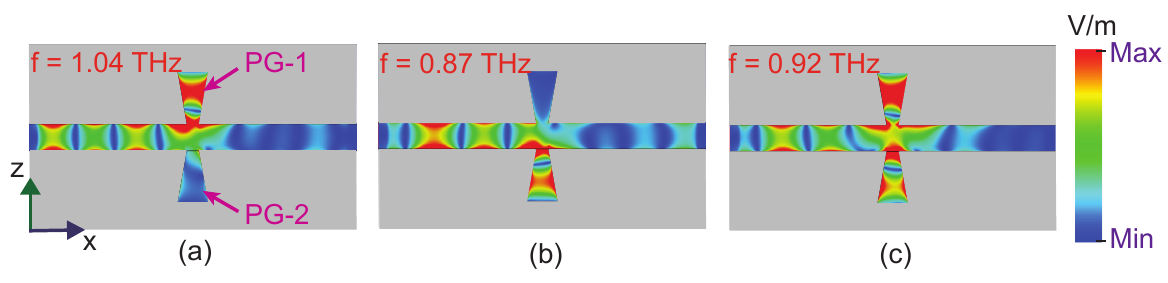}
\caption{(a) – (c) represent electric field profiles in z-x plane supported by PG-1 and PG-2, individually and both PG-1 $\&$ PG-2 together at different frequencies i.e. 1.04 THz, 0.87 THz, and 0.92 THz, respectively. The different parameters of the proposed terahertz waveguide are kept fixed: $l_1 = 50 \mu m$, $l_2 = 120 \mu m$, $w = 800 \mu m$, $h = 200 \mu m$.}\label{fig3}
\end{figure}\\ 
	To get more insight into the PIT-effect through coupling between the two radiative modes supported by PG-1 and PG-2, we examined electric field profiles supported by the waveguide at three different frequencies 1.04 THz, 0.87 THz, and 0.92 THz. From Fig.\ref{fig2}(a) we have confirmed that PG-1 and PG-2 exhibit resonant modes with resonance frequencies 1.04 THz and 0.87 THz, respectively, whereas combined structure of PG-1 and PG-2 exhibit PIT-effect at 0.92 THz. The results of electric field profiles supported by the waveguide configuration in z-x plane are depicted in Fig.\ref{fig3}(a) – (c). From the figures Fig.\ref{fig3}(a) and (b), it can be observed that PG-1 and PG-2 are excited at their corresponding resonant frequencies. In Fig.\ref{fig3}(c), it can be noticed that both pyramidal grooves PG-1 and PG-2 are excited at 0.92 THz, which corresponds to  the frequency of PIT-effect.

\section{Active Modulation of Plasmon Induced Transparency Window}
Next, we investigate the active modulation of the PIT-effect. In order to do so, we varied the refractive index value of the dielectric material in PG-2 from $n$ = 1.1 to 1.3, whereas PG-1 remains empty. \begin{figure}[h!]
\centering\includegraphics[width=13cm]{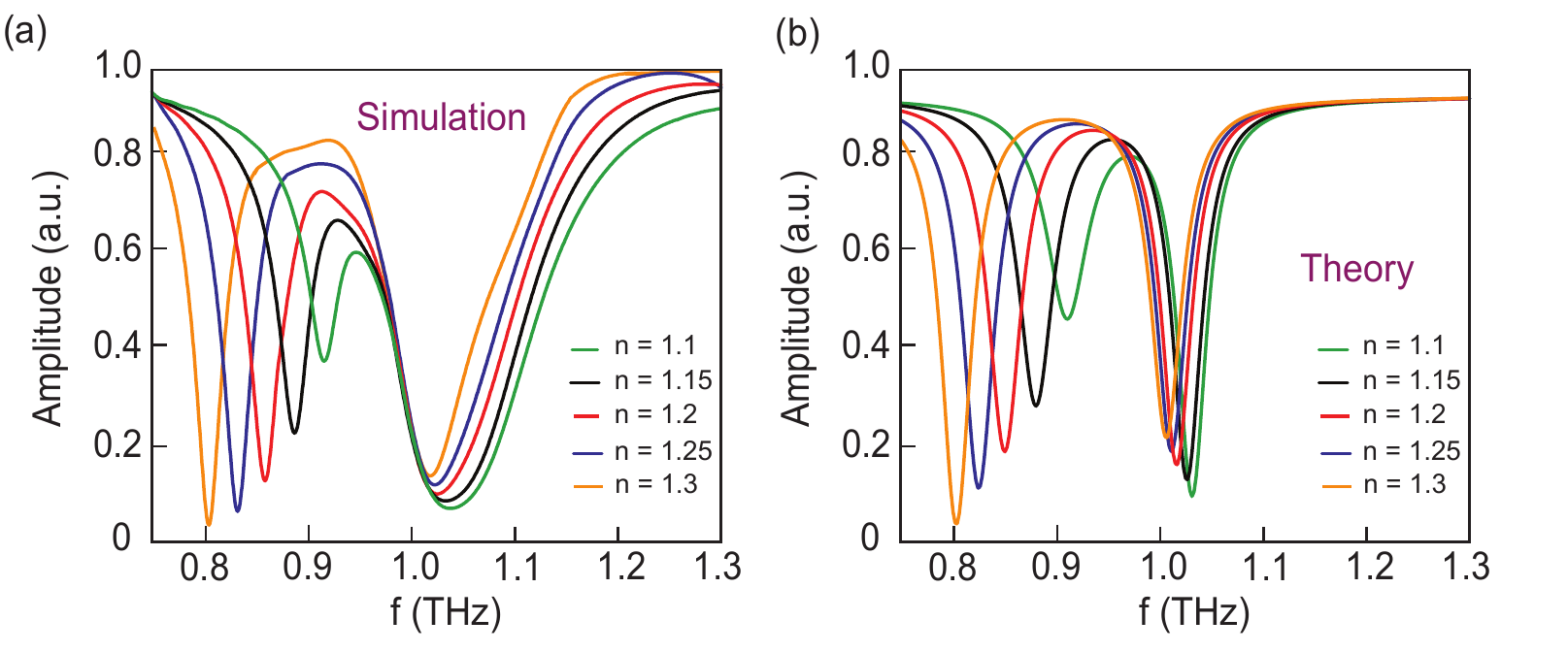}
\caption{Modulation of the PIT window of the proposed terahertz waveguide for different refractive index (n) value of dielectric material in PG-2. Different color traces indicate different $‘n’$ values of dielectric. The other parameters are: $l_1 = 50 \mu m$, $l_2 = 120 \mu m$, $w = 800 \mu m$, $h = 200 \mu m$. (a) Represents numerically simulated results while (b) corresponds to theoretical modeling.}\label{fig4}
\end{figure}The other physical parameters of the waveguide remain fixed: $l_1 = 50 \mu m$, $l_2 = 120 \mu m$, $ w = 800 \mu m$, $h = 200 \mu m$. The corresponding numerically obtained transmission spectra for varying $'n'$ values are shown in Fig.\ref{fig4}(a). It can be observed from the figure that there is a minimal change in the frequency of the resonant mode supported by PG-1, but we observe substantial red shifting in resonance frequency of the mode supported by PG-2. The change in refractive index value causes a red shift along with the broadening of the PIT-window. Hence, we found that the active modulation of PIT-effect can be accomplished with the varying the refractive index of the dielectric material, without causing a change in the physical dimensions of the structures.\\
Next, we again employ the three level plasmonic model to validate numerical findings and understand the coupling mechanism between the two radiative resonant modes supported by PG-1 $\&$ PG-2. The transmission coefficients obtained using theoretical modeling are plotted in Fig.\ref{fig4}(b) for different refractive index values (n) of the dielectric material in PG-2. Different color traces correspond to different values of refractive index (n). Theoretical modeling gives rise to synonymous transmission spectrum as obtained through numerical simulations. The modeling results in same resonance frequencies of the resonant modes along with the PIT-window for varying refractive indices for specific set of fitting parameters given in Table-\ref{tab:table2}. From the table, one can notice that as we increase the refractive index value of dielectric material in PG-2, the coupling coefficient value increases. It is due to the fact that the propagating terahertz surface waves along the gap undergo more interaction as we increase the refractive index value of dielectric material.   
\begin{table}[ht]
	\centering
			\begin{tabular}{|c|c|c|c|c|c|}
			\hline
				R.I. (n) of dielectric & $G$&$\kappa$& $\gamma_a$& $\gamma_b$\\material of PG-2 &(THz)&(THz)&(THz)&(THz)\\
				\hline
				1.1   & 0.3 & 0.03 & 0.108 & 0.155 \\
				1.15   & 0.3 & 0.035 & 0.11 & 0.13 \\
				1.2   & 0.3 & 0.04 & 0.0112 & 0.125 \\
				1.25   & 0.3 & 0.045 & 0.114 & 0.12 \\
				1.3  & 0.3 & 0.05 & 0.116 & 0.115 \\
				\hline
			\end{tabular}
\caption{\label{tab:table2} Parameters for the theoretical fit for various refractive index (n).}		
\end{table}\\

\section{Tunability of Transparency Window Using Silicon Sheet}
\begin{figure}[h!]
\centering\includegraphics[width=13cm]{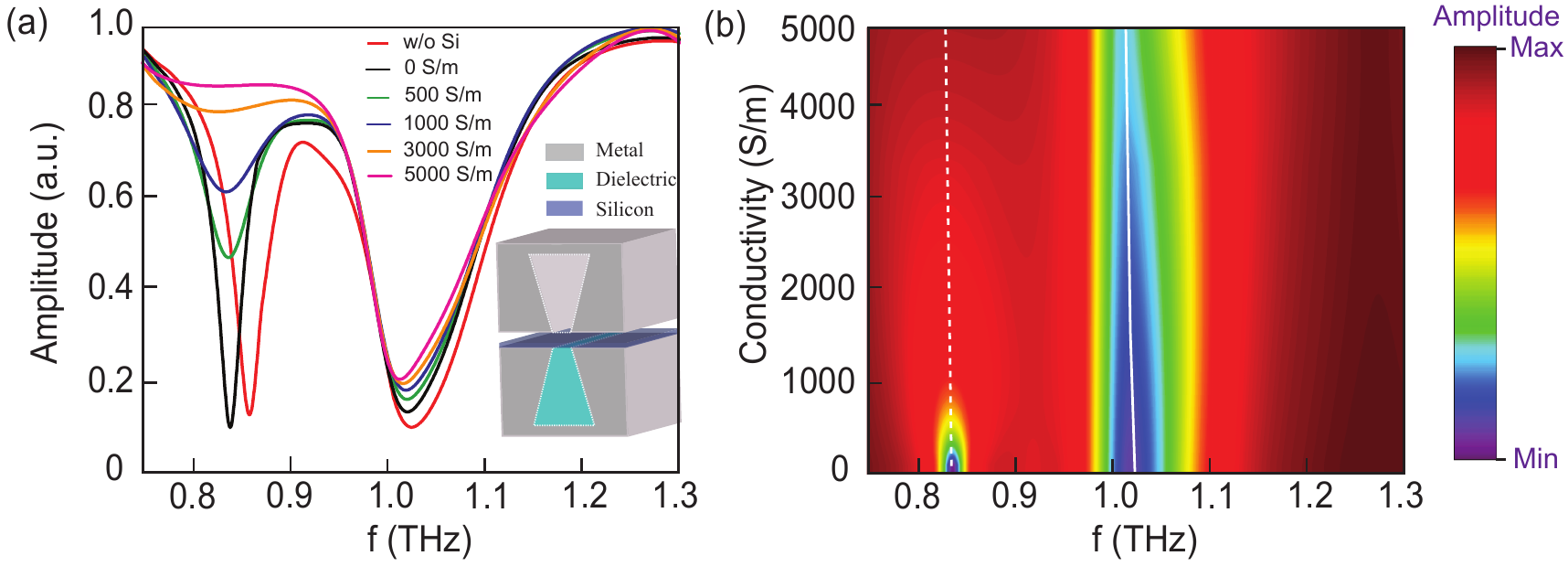}
\caption{(a) Switching and tunability of transparency window for fixed parameters of double parallel groove terahertz waveguide: $l_1 = 50 \mu m$, $l_2 = 120 \mu m$, $w = 800 \mu m$, $h = 200 \mu m$. Numerically obtained transmission spectra of proposed terahertz waveguide for different conductivity of silicon layer placed on the top of PG-2. (b) Contour plot for different conductivity of the silicon layer. Color bar shows the amplitude of transmission signal.}\label{fig5}
\end{figure}

In order to get a better understanding of switching and tunability of the plasmon induced transparency window, we have placed a sheet of semiconducting material like silicon (Si) just above the PG-2 in the gap of the waveguide configuration as shown in the inset of Fig.\ref{fig5}(a).  The silicon material properties are chosen from the experimental data \cite{52}. We have varied the electrical conductivity of the silicon (Si) sheet from 0 $S/m$ to 5000 $S/m$. The corresponding transmission spectra are plotted in Fig.\ref{fig5}(a) for different values of conductivity. Different color traces indicate transmission spectra corresponding to different values of electrical conductivity of silicon sheet. From the plot, it is evident that there is a substantial decrease in the amplitude of the resonant mode supported by PG-2 when conductivity value of silicon sheet is increased. On the other hand, the amplitude of the resonance mode supported by PG-1 remains almost unchanged. It is found that the resonant mode supported by PG-2 completely vanishes at a conductivity value of 5000 S/m. This is owing to the fact that the silicon  sheet over PG-2 becomes highly conducting in nature which forbids the propagating terahertz wave to interact with the pyramidal groove, PG-2. Hence, the resonant mode supported by PG-2 is not observed at high conductivity value.\\\\
Next, we present a comprehensive analysis of switching and tunability of PIT–window through a contour plot of the numerically obtained transmission results for the varying conductivity value of silicon. We have investigated the transmission spectra for forty different electrical conductivity values from 0 S/m to 5000 S/m. The contour plot is depicted in Fig.\ref{fig5}(b). In the figure, the frequency and electrical conductivity values of silicon are shown along the x and y-axis, respectively while the color bar indicates the amplitudes of the transmission spectrum. The solid and dotted white lines the positions of resonant modes supported by PG-1 and PG-2, respectively.  The resonant mode supported by PG-2 vanishes as electrical conductivity value of silicon increases, which can be observed from the contour plot. 
\section{Conclusion}
In conclusion, we demonstrate actively tunable and switchable plasmon induced transparency effect in a parallel plate terahertz waveguide configuration comprising of air-dielectric grooves. The two pyramidal grooves support resonant modes at two different frequencies adjacent to each other. The strong coupling between modes causes the plasmon induced transparency (PIT) effect in the proposed configuration. A three level plasmonic theoretical model is employed to confirm the PIT-effect. The electric field profiles examined at different frequencies in the PIT window clearly depict the PIT-effect. In order to achieve active modulation of the PIT-effect, the refractive index of dielectric material filled in one of the grooves is varied. We observe that the PIT-window gets broadened when the refractive index increases. We have further demonstrated that the PIT effect can be tuned and resonance can be turned ON and OFF by varying the conductivity of silicon sheet placed between the grooves. The proposed study is highly significant wherever an active modulation of PIT effect and resonance switching is required. The information can ultimately be helpful in the construction of active components operating at terahertz frequencies. The proposed concept of active modulation of PIT effect in the terahertz regime can also be pursued in other regions of the electromagnetic spectrum. 

\end{document}